\documentclass[reprint,aps,prx,twocolumn]{revtex4-2}

\usepackage{graphicx}% Include figure files
\usepackage{dcolumn}% Align table columns on decimal point
\usepackage{bm}% bold math

\usepackage{hyperref}
\usepackage{textcomp}
\usepackage{csquotes}

\usepackage{amssymb}
\usepackage{amsthm}
\usepackage{amsfonts}
\usepackage{amssymb}
\usepackage{graphicx}
\usepackage{comment}

\usepackage{subfiles}
\usepackage{subfig}
\usepackage{mathtools}
\usepackage{pgfplots}
\pgfplotsset{/pgf/number format/use comma,compat=newest}
\usepackage{url}
\usepackage{tikz}
\usepackage{notations}
\usepackage{bbm}
\usepackage{floatflt}
\usepackage{subfig}

\newcommand{\E}{\mathbb{E}}
\newcommand{\R}{\mathbb{R}}
\newcommand{\C}{\mathbb{C}}

\newcommand{\D}{\mathcal{D}}

\newtheorem*{result}{Result}
\newtheorem*{algorithm}{Algorithm}

\newenvironment{talign*}
 {\csname align*\endcsname}
 {\endalign}

\begin{document}

\preprint{APS/123-QED}

\title{Information limits and Thouless-Anderson-Palmer equations \\for spiked matrix models with structured noise}
%Bayesian inference for structured spiked models:\\ Information limits, Thouless-Anderson-Palmer equations and an efficient algorithm}
%information limits, Thouless-Anderson-Palmer equations and universality}% Force line breaks with \\

\author{Jean Barbier}
\thanks{All authors contributed equally and names are ordered alphabetically. Correspondence should be sent to \texttt{jbarbier@ictp.it}.}
\author{Francesco Camilli}
%\email{fcamilli@ictp.it}
\author{Yizhou Xu}
%\email{yxu@ictp.it}
\affiliation{
 The Abdus Salam International Centre for Theoretical Physics\\
 Strada Costiera 11, Trieste 34151, Italy
}%


\author{Marco Mondelli}
%\email{marco.mondelli@ist.ac.at}
\affiliation{
 Institute of Science and Technology Austria, \\
    Am Campus 1, 3400 Klosterneuburg, Austria
}%

%\date{\today}

\begin{abstract}
We consider a prototypical problem of Bayesian inference for a structured spiked model: a low-rank signal is corrupted by additive noise. While both information-theoretic and algorithmic limits are well understood when the noise is a Gaussian Wigner matrix, the more realistic case of structured noise still proves to be challenging. To capture the structure while maintaining mathematical tractability, a line of work has focused on rotationally invariant noise. However, existing studies either provide sub-optimal algorithms or which are limited to a special class of noise ensembles. In this paper, using tools from statistical physics (replica method) and random matrix theory (generalized spherical integrals) we establish the first characterization of the information-theoretic limits for a noise matrix drawn from a general trace ensemble. Remarkably, our analysis unveils the asymptotic equivalence between the rotationally invariant model and a surrogate Gaussian one. Finally, we show how to saturate the predicted statistical limits using an efficient algorithm inspired by the theory of adaptive Thouless-Anderson-Palmer (TAP) equations. 
\end{abstract}

\maketitle

%\tableofcontents

\section{Introduction}

Recovering a low-rank signal from a high-dimensional observation corrupted by noise is an ubiquitous problem, appearing e.g.\ in sparse principal component analysis (PCA) \cite{johnstone2009consistency, johnstone2004sparse}, community detection \cite{SBM_abbe18,moore2017computer}, group synchronization \cite{perry2018message} and sub-matrix localization or clustering \cite{lesieur2015mmse}. In this paper, we consider the prototypical task of estimating the rank-1 signal $\bX^*\bX^{*\intercal}\in\mathbb R^{N\times N}$ from a symmetric matrix $\bY$ of noisy observations given by 
\begin{align}\label{eq:model}
\bY=\frac{\lambda}{N}\bX^*\bX^{*\intercal}+\bZ,
\end{align}
where $\lambda\ge 0$ represents the signal-to-noise ratio (SNR) and $\bZ\in \mathbb R^{N\times N}$ is additive noise. This is often referred to as the Johnstone spiked covariance model \cite{johnstone2001distribution}, and it was originally formulated as a probabilistic model for PCA. Starting with the seminal result of \cite{baik2005phase}, the behavior of eigenvalues and eigenvectors of \eqref{eq:model} has been extensively studied in statistics and random matrix theory, see e.g.\ \cite{bai2012sample,baik2006eigenvalues,benaych2011eigenvalues,capitaine2009largest,feral2007largest,knowles2013isotropic}. Specifically, the authors of \cite{baik2005phase} identified a phase transition phenomenon -- named BBP after their initials -- tuned by the SNR $\lambda$: above the transition, the largest eigenvalue of $\bY$ detaches from the bulk of the spectrum containing the noise eigenvalues, and the top eigenvector of $\bY$ is correlated with $\bX$; below the transition, the largest eigenvalue sits at the edge of the bulk, and the top eigenvector exhibits vanishing correlation with $\bX$. %has no correlation with the signal.

Going beyond the estimator obtained from the top eigenvector of $\bY$, 
a line of work has focused on Approximate Message Passing (AMP) algorithms. Originally proposed in the context of compressed sensing \cite{donoho2009message} and CDMA \cite{kabashima2003cdma}, AMP methods have since been developed for numerous high-dimensional inference problems, including the estimation of low-rank matrices \cite{it-PCAMontanari14,montanari2017estimation} as in \eqref{eq:model}, generalized linear regression \cite{rangan2011generalized,sur2019modern} and inference in multi-layer models \cite{manoel2017multi}. The popularity of the AMP paradigm stems from its attractive features: \emph{(i)} AMP can be tailored to take advantage of structural information about the signal, in the form of a Bayesian prior; \emph{(ii)} the AMP performance in the high-dimensional limit is precisely characterized by a low-dimensional deterministic recursion known as state evolution \cite{Bayati_Montanari11,bolthausen2014iterative}; and \emph{(iii)} using state evolution, it has been proved that AMP achieves Bayes-optimal performance in a number of settings \cite{barbier2019optimal,montanari2017estimation} and, even when information-theoretic limits are not met, AMP remains optimal among a vast class of efficient algorithms \cite{celentano2020estimation, montanari2022equivalence}.

However, most theoretical studies on low-rank matrix estimation are limited by an \emph{i.i.d.} (independently and identically distributed) hypothesis on  the noise matrix components. In this setting, the fundamental limits of inference are well understood \cite{2016arXiv161103888L}, and they are achieved by an AMP algorithm, unless there is a statistical-to-computational gap. While some of the results on AMP can be generalized to the broader class of i.i.d.\ sub-Gaussian matrices via universality arguments \cite{bayati2015universality,chen2021universality}, the i.i.d.\ assumption is rather limiting: these models remain structureless, and no concrete correlations can appear in the data matrices. A way to relax the identicality assumption was proposed in the mathematical physics literature of spin glasses, in the context of the Sherrington-Kirkpatrick model. Specifically the authors of \cite{MultiSK_Contucci,Panchenko_multi13}, and later \cite{MSKNL,DBMNL,Bates-Sohn-MSK,Guionnet_inhomogeneous_noise}, consider random couplings whose variances depend on the index labeling the coupled sites. This idea also appeared earlier in the context of inference, under the name of spatial coupling \cite{felstrom1999time,kudekar2011threshold,XXT}. Yet, in the mentioned works the independence hypothesis still remains crucial.

In the seminal papers \cite{Marinari_ParisiI,Marinari_ParisiII,Parisi_Potters}, the authors considered instead a class of \emph{rotationally invariant matrices}, which break the independence between the elements of the coupling matrices, leaving a model which is still tractable. The amount of works in this setting, or similar ones, see for instance \cite{opper2016theory,adatap,maillard2019high} for spin glasses, and \cite{rangan2019vector, gerbelot2020asymptotic, fan2022approximate, zhong2021approximate, mondelli2021pca, venkataramanan2022estimation} in inference, shows a growing interest towards the topic.
Even if the performance of spectral PCA can be predicted with a fairly generic additive rotationally invariant noise (see e.g.\ \cite{benaych2011eigenvalues}), establishing how to factor in also the prior information in the inference procedure, as well as characterizing information-theoretic limits has proven to be significantly more challenging. 

The recent paper \cite{PNAS-Barbier-Camilli} takes a step forward by revealing that, in order to achieve the information-theoretic limits, it is necessary to apply a peculiar pre-processing function to the data $\bY$ that depends on the type of correlations in the noise. %The study is initially carried out only for rotational invariant noise matrices, but numerical evidence supports the validity of the derivation also for other types of noises, whose eigenvectors are extracted by covariance matrices of real data. 
Despite the new mechanism pinpointed by \cite{PNAS-Barbier-Camilli}, the analysis has remained limited only to certain classes of noise distributions, until now. The goal of the present paper is precisely to elaborate a concise theory, and to formulate implementable algorithms, that can be tailored to treat \emph{any} kind of rotationally invariant noise coming from a trace ensemble, i.e., whose distribution is encoded in a matrix potential.

\subsection{Our contributions}
\begin{itemize}
    \item[\emph{i)}] Using the celebrated \emph{replica method} \cite{mezard1990spin} %, combined with some basic complex analysis tools, 
    and the inhomogeneous spherical integral of \cite{PNAS-Barbier-Camilli}, we %were able to 
    compute the information-theoretic limits for low-rank estimation in the presence of rotationally invariant noise. Specifically, we consider the  teacher-student scenario in \eqref{eq:model}, i.e., the teacher plants a rank-1 spike matrix inside an extensive-rank noise bulk, and we compute the mutual information between the observation $\bY$, and the planted signal $\bX^*\bX^{*\intercal}$. 
    \begin{itemize}
        \item[\emph{i-a)}] We simplify and solve numerically the fixed point equations coming from the replica symmetric variational potential for the mutual information. Remarkably, thanks to some inherent symmetries of the model, called Nishimori identities, the fixed point equations, which have functions among their unknowns, reduce to one simple scalar equation.
        \item[\emph{i-b)}] The noise not being Gaussian, one cannot use the usual \emph{I-MMSE} relation \cite{guo2005mutual} to compute the minimum mean-square error (MMSE) from the mutual information. Nevertheless, our final formula for the mutual information consists in a variational principle, whose order parameters, at their stationary values, yield the MMSE when properly combined.
    \end{itemize}

    \item[\emph{ii)}] We also express the mutual information between data and ground truth using the AdaTAP formalism of  \cite{adatap}, relying on the validity of the latter in the presence of a spike. This approach outputs the pre-processing function as a functional of the matrix potential of the noise ensemble. {Moreover, the stationary point of the AdaTAP mutual information yields a set of fixed point equations, the TAP equations, that we turn into an inference algorithm for the spike with an appropriate update rule.}

    \item[\emph{iii)}] We run numerical experiments supporting the consistency between the fixed point of the TAP equations of point \emph{ii)} and our replica prediction for the MMSE.
  {We test our TAP-inspired algorithm also on covariance matrices built from a bioinformatics dataset (Hapmap3 \cite{international2010integrating}), where we plant suitably generated spikes. We verify that, even if the noise matrix is not rotational invariant, our TAP algorithm, when properly initialized, is still able to find the spikes with a performance close to the one predicted by the Bayes-optimal replica theory. This evidence supports the conjecture already formulated in \cite{PNAS-Barbier-Camilli}, stemmed from \cite{dudeja2022universality2,dudeja2022universality}, about a possible universality of our information-theoretic analysis, which could thus extend beyond the rotational invariance assumption to cases where the eigenbasis of the noise is invariant under more restrictive transformations.}
\end{itemize}

\textbf{Code repository:} All the codes used for this paper are available at \footnote{Github repository at \href{ https://github.com/xu-yz19/spiked-matrix-models-with-structured-noise}{https://github.com/xu-yz19/spiked-matrix-models-with-structured-noise}}.

\section{Setting}
Let the data be constructed, conditionally on the unknown spike $\bX^*\bX^{*\intercal}$, according to \eqref{eq:model},
with a randomly generated noise matrix
\begin{align}
 \bZ&=\bO^\intercal \bD \bO\sim C_V\exp\Big(-\frac N2\text{Tr}V(\bZ)\Big)\,d\bZ,
\end{align}
where $d\bZ=\prod_{i\leq j}^N dZ_{ij}$ and $Z_{ij}=Z_{ji}$ for any $1\leq i,j\leq N$, and $C_V$ is a normalizing constant depending on the matrix potential $V$.
The matrices have $O(1)$ spectral norm (i.e., independent on $N$) and the signal vector $\bX^*$ has i.i.d.\ entries drawn from a prior distribution $P_X$ with zero mean and unit second moment. With an abuse of notation, we use $P_X(\bx)$ instead of $P_X^{\otimes N}(\bx)$. Here, and throughout the paper, any function $f$ applied to a symmetric matrix $\bM$ with eigendecomposition $\bM=\bO\bD\bO^\intercal$ is actually applied only to its eigenvalues $(D_i)_{i\leq N}$: $f(\bM):=\bO f(\bD)\bO^\intercal$, $f(\bD)=\text{diag}(f(D_i))_{i\leq N}$. 

We highlight that, compared to the previous work \cite{PNAS-Barbier-Camilli} where the matrix potential was restricted to be a low degree polynomial, here $V$ can be \emph{any} analytic function.

The posterior measure of the problem is given by
\begin{equation}
    dP_{X|Y}(\bx|\bY)=\frac{C_V}{P_Y(\bY)}dP_X(\bx)e^{-\frac{N}{2}\text{Tr}V(\bY-\frac{\lambda}{N}\bx\bx^\intercal)},
\end{equation}
where the evidence, i.e., the probability of the observations, is
\begin{equation}
     P_Y(\bY)=C_V\int_{\mathbb{R}^N} dP_X(\bx)e^{-\frac{N}{2}\text{Tr}V(\bY-\frac{\lambda}{N}\bx\bx^\intercal)}.
\end{equation}
Our main object of interest is the free entropy (or minus the free energy) defined as
\begin{align}
    f_N:=\frac{1}{N}\E_\bY\log \mathcal{Z}(\bY),
\end{align}
where 
\begin{align}\label{eq:partition-function}
    \mathcal{Z}(\bY):=\int_{\mathbb{R}^N} dP_X(\bx)e^{-\frac{N}{2}\big(\text{Tr}V(\bY-\frac{\lambda}{N}\bx\bx^\intercal)-\Tr V(\bZ)\big)}
\end{align}
and in particular its high-dimensional limit $f=\lim_{N\to\infty} f_N$. From the above, we can define a Hamiltonian function $\mathcal{H}_N(\bx,\lambda,\bX^*,\bZ)=\mathcal{H}_N(\bx)$ equal to
\begin{align}
    \mathcal{H}_N(\bx)=\frac{N}{2}\big(\Tr V(\bY-\frac{\lambda}{N}\bx\bx^\intercal)-\Tr V(\bZ)\big)
\end{align}
and interpret the problem as a spin glass model where $\bY$, or equivalently the independent $\bZ$ and $\bX^*$, play the role of quenched disorder. The subtraction of the term $\Tr V(\bZ)$ is needed to ensure that the Hamiltonian remains of $O(N)$, and thus the free entropy stays $O(1)$ in the thermodynamic limit $N\to\infty$. In fact,
\begin{align}
    \mathcal{H}_N(\bx)&=\frac{\lambda}{2}\int_0^1 dt \,\Tr \big[V'\big(\bZ+\frac{t\lambda
    }{N}(\bX^*\bX^{*\intercal}-\bx\bx^\intercal)\big)\nonumber\\
    &\qquad\times\big(\bX^*\bX^{*\intercal}-\bx\bx^\intercal\big)\big]=O(N),
\end{align}
since the difference between the two projectors in the last line is at most rank two and the related eigenvalues remain $O(1)$.

The free entropy is intimately connected to the mutual information between data and signal:
\begin{align}
    I(\bX^*\bX^{*\intercal};\bY)&=-\EE_\bY \log P_Y(\bY)+\EE_\bZ \log C_Ve^{-\frac{N}{2}\Tr V(\bZ)}\nonumber\\
    &=-Nf_N.
\end{align}

Finally, let us recall some basic concepts from random matrix theory. For a square symmetric random matrix $\bM\in\mathbb{R}^{N\times N}$ with real eigenvalues $\{\lambda_i(\bM)\}$, denote the associated resolvent matrix
\begin{equation}
    \bG_M(z):=(zI_N-\bM)^{-1}, \quad z\in\mathbb{C}\,\backslash\, \{\lambda_i(\bM)\}_{i\le N}.
\end{equation}
Notice that the resolvent matrix shares the same eigenvectors with $\bM$. Assume that the empirical spectral density (ESD) $\hat{\rho}_M^{(N)}$ of $\bM$ converges weakly almost surely to a distribution $\rho_M$, i.e.
\begin{align}
    \hat{\rho}_M^{(N)}=\frac{1}{N}\sum_{i=1}^N\delta_{\lambda_i(\bM)}\xrightarrow[w.a.s.]{N\to\infty}\rho_M.
\end{align}
We can then define the Stieltjes transform associated with the random matrix ensemble of $\bM$:
\begin{equation}
    g_M(z):=\E_D\frac{1}{z-D},\quad D\sim\rho_M.
\end{equation}
$g_M(z)$ is well defined for $z\in\mathbb{C}$ outside the support of $\rho_M$. Under the aforementioned hypothesis on the ESD, the Stieltjes transform is closely related to the resolvent matrix through
\begin{equation}
    \lim_{N\to\infty}\frac{1}{N}\text{Tr}\bG_M(z)=g_M(z)
\end{equation}
almost surely. Denote the inverse function of $g_M(z)$ as $\zeta_M(g)$. Then the R-transform of $\bM$ is given by
\begin{equation}
    \mathcal{R}_\bM(g):=\zeta_M(g)-\frac{1}{g}.
\end{equation}
The resolvent and the R-transform play a crucial role in our analysis since they encode all the relevant combinatorics of the random matrix ensemble. As we shall see later, the resolvent allows us to define a new family of order parameters, which are quadratic forms of replicas drawn from the posterior with the resolvent mediating their product. In one shot, this new family of overlaps encompasses all the new order parameters that were introduced in \cite{PNAS-Barbier-Camilli}, as well as additional ones.

\section{Information limits through \\the replica method}
In this section we analyse the spiked model with generic rotationally invariant noise using the powerful replica method from statistical physics of disordered systems \cite{mezard1990spin,mezard2009information}. This method is non-rigorous, but it is believed to be exact for the asymptotic analysis of a broad class of spin glass, inference and learning models. Historically, one of the first proofs of its exactness was given for the Sherrington-Kirkpatrick model by Guerra \cite{guerra2003broken} and Talagrand \cite{talagrand2006parisi}, and later remarkably refined by Panchenko leveraging ultrametricity \cite{panchenko2013parisi}. Moreover, the replica symmetry assumption we are going to employ during the analysis is intimately connected to concentration-of-measure phenomena proven in broad generality in optimal Bayesian inference \cite{barbier_strong_CMP,barbier2019overlap}. We therefore conjecture that the analysis below leads to asymptotically exact formulas. For further discussions on the topic, we refer the reader to \cite{barbier2016proof,el2018estimation,BarbierMacris2019,krzakala_limits20,Barbier_IMAIAI21}.

We now state our main result from the information-theoretic perspective. This comes in the form of a variational formula for the free entropy.
The physical meaning of some of the order parameters entering these formulas is given in the replica analysis of the next section.

\newpage
\begin{result}[Information-theoretic limits: Replica free entropy and minimum mean-square error] 
\begin{widetext}
Let $\Gamma$ be an arbitrary contour in the complex plane $\C$ that encircles all eigenvalues of the matrix $\bY-\bx\bx^\intercal/N$ for any choice of $\bx$ with positive measure according to the prior $P_X$. %(any rectangle encircling all $\R$ is fine).
Let $Z\sim \mathcal{N}(0,1)$ and $X\sim P_X$ as well as $D,D'\sim \rho_Z$ i.i.d.\ from the noise asymptotic spectral density. The replica free entropy at the replica symmetric level (which is exact in Bayes-optimal inference) is given by
\begin{align}
f&= {\rm extr}\Big\{\frac{1}{4\pi i}\oint_\Gamma dz\Big[V'(z)\log(1+\lambda B(z))-2\frac{\lambda\hat{B}(z)M(z)^2}{1-\lambda g_Z(z)}+2\hat{M}(z)M(z)+2\hat{B}(z)B(z)\Big]\nonumber \\
&\qquad+\frac{v\hat{v}}{2}-\hat{m}m+\frac{q\hat{q}}{2}+\mathbb{E}_{Z,X}\log\int_{\mathbb{R}} dP_X(x)\exp\Big(\sqrt{\hat{q}}Zx-\frac{\hat{q}+\hat{v}}{2}x^2+\hat{m}Xx\Big)+m\bar{m}+\frac{v\bar{v}}{2}-\frac{q\bar{q}}{2}-\frac{1}{2}\nonumber \\
&\qquad-\frac{1}{2}\mathbb{E}_D\log(\bar{v}-\bar{q}+2\tilde{B}(D))-\frac{1}{2}\mathbb{E}_D\frac{\bar{q}-(\bar{q}+\tilde{M}(D))^2}{\bar{v}-\bar{q}+2\tilde{B}(D)}-\frac{1}{2}\log(v-q)-\frac{q-m^2}{2(v-q)}\Big\}\nonumber\\
&\qquad+\frac{1}{4\pi i}\oint_\Gamma dz V'(z)\log(1-\lambda g_Z(z)),
\label{eq:variation_free_energy}
\end{align}
with an extremization w.r.t.\ nine scalar order parameters $(m,q,v,\hat{m},\hat{q},\hat{v},\bar{m}, \bar{q}, \bar{v})$ and four functions $(M, B, \hat{M}, \hat{B})$ from $\mathbb{C}$ to $\mathbb{C}$. The extremization selects the solution of the saddle point equations, obtained by equating to zero the gradient of the replica potential $\{\cdots\}$, which maximizes it. After simplifications of the replica saddle point equations, this can also be written as 
\begin{align}
    &f=\max_{\mathcal{M}^{\rm RS}}f^{\rm RS}(m,\hat{m})-\frac{\lambda }{2}\mathbb{E}_DV'(D) , \label{f=maxfrs}
    \\
    &f^{\rm RS}(m,\hat{m}):=-\frac{\lambda^2}{2}\mathbb{E}_{D,D'}Q(D)Q(D')H(D)H(D')\frac{V'(D)-V'(D')}{D-D'}-\frac{m^2}{2(1-m)}-\frac{1}{2}\log(1-m)-\frac{m}{2}\nonumber\\
    &\quad+\mathbb{E}_{Z,X}\log\int_{\mathbb{R}} dP_X(x)\exp\Big(\sqrt{\hat{m}}Zx-\frac{\hat{m}}{2}x^2+\hat{m}Xx\Big)+\frac{1}{2}\mathbb{E}_D\log H(D)-\frac{1}{2}\mathbb{E}_D\Big(\hat{m}-\frac{1}{1-m}-Q(D)^2\Big)H(D),\label{eq:simiplified_f}
\end{align}
with
\begin{align}
H(x)&:=\Big(\frac{1}{1-m}-\hat{m}-J(x)\Big)^{-1},\label{eq:H}\\
J(x)&:=\lambda V'(x)-\lambda^2\mathbb{E}_{D\sim \rho_Z}\frac{V'(x)-V'(D)}{x-D},\label{eq:J}
\end{align}
and $Q(x)$ is the solution of
\begin{equation}
    Q(x)=\hat{m}-\frac{1}{1-m}+\lambda^2\mathbb{E}_D\frac{V'(x)-V'(D)}{x-D}Q(D)H(D).\label{eq:Q}
\end{equation}
Finally, $\mathcal{M}^{\rm RS}$ represents the set of solution(s) of the following fixed point equations:
\begin{align}
    \begin{cases}
        \hat{m}=-\mathcal{R}_{J(\bZ)} (1-m),\\
    m=\mathbb{E}_{Z,X}X\langle x\rangle_{\hat{m}},
    \end{cases}
    \label{eq:saddle_point1}
\end{align}
where $\langle\,\cdot\,\rangle_{\hat{m}}$ denotes the expectation w.r.t. the posterior of a scalar Gaussian channel with signal-to-noise ratio $\hat m$:
\begin{equation}    \langle f(x)\rangle_{\hat{m}}= \langle f(x)\rangle_{\hat{m}}(Z,X):=\frac{\int dP_X(x)e^{\sqrt{\hat{m}}Zx+\hat{m}xX-\frac{\hat{m}}{2}x^2}f(x)}{\int dP_X(x)e^{\sqrt{\hat{m}}Zx+\hat{m}xX-\frac{\hat{m}}{2}x^2}}.\label{eq:posterior}
\end{equation}

Recall that $\int dP_X(x)x^2=1$. Let $m_*$ be the value of the order parameter $m$ picked by the above extremization, i.e., the solution of \eqref{eq:saddle_point1} which maximizes $f^{\rm RS}(m,\hat{m})$. The asymptotic minimum mean-square error corresponding to the %spike
Bayes-optimal estimator (i.e., the posterior mean) $\E[\bX^{*}\bX^{*\intercal}\mid \bY]$ is given by
\begin{align}    \lim_{N\to\infty}\frac1{N^2}\E\big\|\bX^{*}\bX^{*\intercal}-\E[\bX^{*}\bX^{*\intercal}\mid \bY]\big\|_{\rm F}^2=1 - m_*^2.
\end{align}
\end{widetext}
\end{result}

The above generic formula for the replica free entropy is rather involved. Note that, fortunately, it is not needed to evaluate it in order to access the most interesting quantity, namely, the MMSE. The latter is instead obtained by solving the much simpler system \eqref{eq:saddle_point1}. The necessary simplifications to go from the replica free entropy to these simpler equations will be explained in sec.~\ref{sec:simplifications}.

\subsection{Analysis by the replica method}
We now provide the derivation of the previous result. Before replicating the partition function we are going to re-express it in a more amenable form. We start by extracting the matrix entering the potential in the log-likelihood term of the partition function using Cauchy's formula. We will then repeatedly use Sherman-Morrison's formula to deal with inverses of rank-one perturbations of matrices: 
%\begin{widetext}
\begin{align}
\mathcal{Z}&=\int_{\R^N}dP_X(\bx)e^{-\frac{N}{2}\text{Tr}\big(V\big(\bY-\frac{\lambda}{N}\bx\bx^\intercal\big)-\Tr V(\bZ)\big)}\label{eq:partition} \\
&=\E_\bx e^{\frac{N}{2}\Tr V(\bZ)-\frac{N}{4\pi i}\text{Tr}\oint_{\Gamma} dzV(z)\big(zI_N-\bY+\frac{\lambda}{N}\bx\bx^\intercal\big)^{-1}}\nonumber\\
&=\E_\bx\exp\Big(\frac{N}{2}\Tr V(\bZ)-\frac{N}{4\pi i}\text{Tr}\oint_\Gamma dzV(z)\nonumber\\
&\qquad\times\Big[\bG_Y(z)-\frac{\lambda}{N}\frac{\bG_Y(z)\bx\bx^\intercal \bG_Y(z)}{1+\frac{\lambda}{N}\bx^\intercal \bG_Y(z)\bx}\Big]\Big)\nonumber\\
&=C\,\E_\bx\exp\Big(\frac{N}{4\pi i}\oint_\Gamma dzV(z)\frac{\lambda}{N}\frac{\bx^\intercal \bG_Y(z)^2\bx}{1+\frac{\lambda}{N}\bx^\intercal \bG_Y(z)\bx}\Big)\nonumber\\
&=C\,\E_\bx\exp\Big(\frac{N}{4\pi i}\oint_\Gamma dzV'(z)\log\Big(1+\frac{\lambda}{N}\bx^\intercal \bG_Y(z)\bx\Big)\Big)\nonumber\\
&=: C\, \Omega(\bY)
\end{align}
%\end{widetext}
where we used $\partial_z \bG_Y(z)=-\bG_Y(z)^2$ and an integration by part in the last equality. Here $$C:=\exp\Big(-\frac{N}{2}\big(\text{Tr}V(\bY)-\Tr V(\bZ)\big)\Big)$$ is a multiplicative constant yielding an additive constant in the free entropy 
\begin{align}\label{2terms}
\frac1N \EE \log \mathcal{Z}=\frac1N\EE \log C+\frac1N \EE \ln \Omega.    
\end{align}
We will compute it separately but for now we focus on the computation of the latter term by the replica method.

We are now ready to replicate. We denote with the replica index 0 the signal $\bX^*=\bx_0$. We then get that 
\begin{align*}
\E \Omega^n&= \EE \int \prod_{a=0}^n dP_X(\bx_a)\\
&\ \ \times\exp\Big(\frac{N}{4\pi i}\sum_{a=1}^n\oint_\Gamma dz V'(z)\log(1+\lambda B^{aa}(z))\Big),
\end{align*}
where we introduce the following order parameters for $1\leq a\leq n$, which are generalized data-dependent self-overlap functions, 
\begin{equation}
    B^{aa}(z):=\frac{1}{N}\bx_a^\intercal \bG_Y(z)\bx_a\in\C.
\end{equation}
As the expectation w.r.t. the signal (replica $0$) is now explicit, the remaining disorder expectation $\EE_{\bO}$ is w.r.t. to the noise Haar distributed eigenbasis only. Indeed, averaging or not w.r.t. to the eigenvalues of the noise does not change the final result as long as its empirical spectral law converges. We can thus consider the $N$-dependent sequence of eigenvalues deterministic. We now introduce delta functions in their Fourier form, together with the conjugate order parameters, in order to fix the $B^{aa}(z)$ definitions. Jointly denote $\D[i\hat{B},B]:=\prod_{a= 1}^n\D[i\hat{B}^{aa},B^{aa}]$ for the differential element in functional (path) integrals. The replicated partition function $\E \Omega^n$ then becomes
\begin{align*}
&\int \D[i\hat{B},B] \prod_{a=0}^ndP_X(\bx_a) \\
&\ \times\exp \Big(\frac{N}{4\pi i}\sum_{a\leq n}\oint_\Gamma dzV'(z)\log(1+\lambda B^{aa}(z))\Big)\\
&\ \times\mathbb{E}_\bO\exp\Big(\sum_{a\leq n}\oint_\Gamma dz\hat{B}^{aa}(z)(NB^{aa}(z)-\bx_a^\intercal \bG_Y(z)\bx_a)\Big).
\end{align*}
The $z$-integral is on the contour $\Gamma$. In order to perform the quenched average over $\bO$  we need to decompose explicitly the data into signal plus noise. The last term can then be simplified using Sherman-Morrison again:
\begin{align}
\frac{\bx_a^\intercal \bG_Y(z)\bx_a}{N}\!&=\!\frac{1}{N}\bx_a^\intercal\Big(\bG_Z(z)+\frac{\lambda \bG_Z(z)\bx_0\bx_0^\intercal \bG_Z(z)}{N(1-\frac{\lambda}{N}\bx_0^\intercal \bG_Z(z)\bx_0)}\Big)\bx_a \nonumber\\
&\!=\!\frac{\bx_a^\intercal \bG_Z(z)\bx_a}{N}+\frac{\lambda M^{a0}(z)^2}{1-\lambda g_Z(z)},
\end{align}
where we introduce the main order parameters, i.e., generalized overlaps between replicas and the ground-truth (which, by Bayes-optimality, also corresponds to the generalized overlap between different replicas): for $1\leq a\leq n$,
\begin{equation}
    M^{a0}(z):=\frac{1}{N}\bx_a^\intercal \bG_Z(z)\bx_0\in\C.
\end{equation}
Note that by definition and independence of signal and noise we have 
\begin{align}
\lim_{N\to\infty}\frac{1}{N}\bx_0^\intercal \bG_Z(z)\bx_0= g_Z(z).    
\end{align}
Let us make a remark concerning the generalized overlap function $M^{a0}(z)$. By expanding in series the resolvent around $z\to +\infty$, we realize that it corresponds to the generating function for an infinite family of scalar overlaps $(\frac{1}{N}\bx_a^\intercal \bZ^k\bx_0)_{k\ge 0}$. A similar observation can be made for $B^{aa}(z)$ which encodes $(\frac{1}{N}\bx_a^\intercal \bZ^k\bx_a)_{k\ge 0}$. The first few of these overlaps are the order parameters identified in \cite{PNAS-Barbier-Camilli}. We note that %hat prevented 
the analysis of \cite{PNAS-Barbier-Camilli} is restricted to 
%to go beyond simple cases of 
low-degree polynomials for the potential $V$. This is because the number of order parameters -- and, therefore, the number of replica saddle point equations -- grows with the degree of the polynomial, which % Thus, this %  is that the number of such order parameters grows with its degree and the number and complexity of the replica saddle point equations too, which 
quickly leads to intractable, non interpretable, formulas. However, by  identifying %cation of 
these generating function order parameters, we can easily encode such infinite families of scalars and write down compact equations. This is one key mechanism that allows us to treat generic potential functions $V$. Similar ideas have been used to study gradient-flow dynamics in \cite{bodin2021rank,bodin2021model,bodin2024random,bodin2022gradient}.

We consider a replica symmetric ansatz: for all replica indices $1\leq a\leq n$, we set
\begin{equation*}
{\rm RS\ ansatz}\mbox{:}\quad 
    \begin{cases}
        (B^{aa}(z),\hat{B}^{aa}(z))=(B(z),\hat{B}(z)),\\
        (M^{a0}(z),\hat{M}^{a0}(z))=(M(z),\hat{M}(z)).
    \end{cases}
\end{equation*}
This implies the following simplifications for the replicated partition function: letting this time $\D[\cdots]=\D[i\hat{B},B,i\hat{M},M]$, we have
%\begin{widetext}
\begin{align}
    \E \Omega^n&=\int \D[\cdots]\exp\Big(\frac{Nn}{4\pi i}\oint_\Gamma dzV'(z)\log(1+\lambda B(z))\nonumber \\
    &\qquad-Nn\oint_\Gamma dz\frac{\lambda\hat{B}(z)M(z)^2}{1-\lambda g_Z(z)}+Nn\oint_\Gamma dz\hat{M}(z)M(z)\nonumber\\
    &\qquad+Nn\oint_\Gamma dz\hat{B}(z)B(z)+NnI^{\rm RS}\Big),
\end{align}
where we also define
\begin{align}  \exp(NnI^{\rm RS})&:=\mathbb{E}_\bO\exp\Big(-\sum_{a\leq n}\oint_\Gamma dz\Big(\hat{B}(z)\bx_a^\intercal \bG_Z(z)\bx_a\nonumber\\
&\qquad+\hat{M}(z)\bx_0^\intercal \bG_Z(z)\bx_a\Big)\Big).
\end{align}
%\end{widetext}
The above integral $I^{\rm RS}$ is an instance of the \emph{inhomogeneous spherical integral} defined and analysed in \cite{PNAS-Barbier-Camilli}. A key property of this integral is that it depends on the replicas only through their overlap structure which, under a replica symmetric ansatz, reads
    \begin{align}\label{RSoverlap}
    {\rm RS\ ansatz}\mbox{:}\quad 
    \begin{cases}
        \bx_a^\intercal \bx_b/N=q,\quad 1\leq a<b\leq n,\\
        \bx_0^\intercal \bx_a/N=m,\quad 1\leq a\leq n,\\
        \bx_a^\intercal \bx_a/N=v,\quad 1\leq a\leq n.
        \end{cases}
    \end{align}
Let us define
\begin{align*}
\tilde{M}(x)&:=\frac{1}{2\pi i}\oint_\Gamma\frac{\hat{M}(z)dz}{z-x},\quad
\tilde{B}(x):=\frac{1}{2\pi i}\oint_\Gamma\frac{\hat{B}(z)dz}{z-x}.
\end{align*}
It is important to note that in general $\tilde{B}(x)\neq\hat{B}(x)$ and $\tilde{M}(x)\neq\hat{M}(x)$ because they might not be holomorphic. Then, the result of the inhomogeneous spherical integral reads \cite{PNAS-Barbier-Camilli}
%\begin{widetext}
\begin{align}
&I^{\rm RS}(q,m,v,\Tilde{B},\tilde{M})=\text{extr}_{(\bar{m},\bar{v},\bar{q})}\Big\{m\bar{m}+\frac{v\bar{v}}{2}-\frac{q\bar{q}}{2}\nonumber\\
&\quad-\frac{1}{2}\mathbb{E}_D\log(\bar{v}-\bar{q}+2\tilde{B}(D))-\frac{1}{2}\mathbb{E}_D\frac{\bar{q}-(\bar{m}+\tilde{M}(D))^2}{\bar{v}-\bar{q}+2\tilde{B}(D)}\Big\}\nonumber\\
&\quad-\frac{1}{2}-\frac{1}{2}\log(v-q)-\frac{q-m^2}{2(v-q)}+O(n).
\end{align}
As before, the extremization picks the maximizing saddle point.
Now, fixing the overlap definitions \eqref{RSoverlap} using additional delta functions in Fourier form, under the same replica ansatz for the Fourier conjugates, we reach after standard manipulations (see, e.g., \cite{PNAS-Barbier-Camilli}):
\begin{widetext}
    \begin{align}
    \E \Omega^n&=\int \D[\cdots]\int dqd\hat q \,dv d\hat v\, dmd\hat m\exp\Big(\frac{Nn}{4\pi i}\oint_\Gamma dzV'(z)\log(1+\lambda B(z))-Nn\oint_\Gamma\frac{\lambda\hat{B}(z)M(z)^2}{1-\lambda g_Z(z)}\nonumber\\
    &\qquad+Nn\oint_\Gamma dz\hat{M}(z)M(z)+Nn\oint_\Gamma dz\hat{B}(z)B(z)+NnI^{RS}(q,m,v,\Tilde{B},\tilde{M})+\frac{Nn}{2}v\hat{v}-Nnm\hat{m}\nonumber\\
    &\qquad-\frac{Nn(n-1)}{2}q\hat{q}+N\log \int_{\R^{n+1}}\prod_{a=0}^ndP_X(x_a)\exp\Big(-\frac{\hat{v}}{2}\sum_{a=1}^n x_a^2+\hat{m}\sum_{a=1}^nx_0x_a+\hat{q}\sum_{1\le a<b\le n}x_ax_b\Big)\Big).
    \end{align}
\end{widetext}
In order to decouple the replicas in the last term we use an Hubbard-Stratonovitch transform: with $Z\sim \mathcal{N}(0,1)$,
\begin{align*}
    &\E_{(x_a)}e^{-\frac{\hat{v}}{2}\sum_{a=1}^n x_a^2+\hat{m}\sum_{a=1}^nx_0x_a+\hat{q}\sum_{1\le a<b\le n}x_ax_b}\\
    &\quad=\E_{Z,x_0}\big(\E_{x}e^{-\frac{\hat{v}+\hat q}{2} x^2+\hat{m}x_0x+\sqrt{\hat{q}} Z x}\big)^n.
\end{align*}
The final steps are then an integration with respect to the order parameters by saddle point, followed by an application of the replica trick (assuming commutation of thermodynamic and replica limits for the saddle point integration),
\begin{align}
\lim_{N\to\infty}\frac1N\E\ln \Omega&=\lim_{N\to\infty}\lim_{n\to 0}\frac1{Nn}\ln\E \Omega^n\nonumber\\
&=\lim_{n\to 0}\lim_{N\to\infty}\frac1{Nn}\ln\E \Omega^n,
\end{align}
as well as a change of variables $(2\pi i\hat{M}, 2\pi i\hat{B})\to(\hat{M},\hat{B})$.

The average limiting free entropy expression still requires us to compute $\frac1N\EE \ln C$. One may think it is irrelevant, however it is a function of the SNR and concurs to the value of the mutual information and some of its fundamental properties (for instance, monotonicity and concavity). We can again use Cauchy's integral representation and the Shermann-Morrison formula for $V(\bY)$:
\begin{align*}
    \frac{1 }{N}\EE\log C&=-\frac{1}{2}\EE\Tr\big(V(\bY)-V(\bZ)\big)\nonumber\\
    &=-\frac{1}{4\pi i}\EE\Tr\oint_\Gamma dz V(z)\big(\bG_Y(z)-\bG_Z(z)\big)\nonumber\\
    &=-\frac{1}{4\pi i}\EE\oint_\Gamma dz V(z)\frac{\lambda}{N} \frac{\bx_0^\intercal \bG^2_Z(z)\bx_0}{(1-\frac{\lambda}{N}\bx_0^\intercal \bG_Z(z)\bx_0)}, 
\end{align*}
which in the large $N$ limit converges to
\begin{align}\label{eq:C_final}
    \frac{1 }{N}\EE\log C&\to \frac{1}{4\pi i} \oint_\Gamma dz V(z) \frac{\lambda \partial _zg_Z(z)}{(1-\lambda g_Z(z))}\nonumber\\
    &=\frac{1}{4\pi i} \oint_\Gamma dz V'(z) \log(1-\lambda g_Z(z))\,. 
\end{align}
Combining everything in \eqref{2terms} yields formula \eqref{eq:variation_free_energy}.

\subsection{Simplifying the saddle point equations} \label{sec:simplifications}
In this section we show how to go from the variational formulation \eqref{eq:variation_free_energy} for the free entropy, to a simpler formula \eqref{f=maxfrs},
\eqref{eq:simiplified_f} with only two order parameters. Before giving the complete set of saddle point equations derived from \eqref{eq:variation_free_energy}, we stress that the physical meaning of some order parameters makes it possible to fix directly their values to their expectation (assuming concentration), obtainable using the Nishimori identities, see \cite[Proposition 15]{lelarge2019fundamental} for a proof.\\

\noindent {\bf Nishimori identity.}\ \ For any bounded function $f$ of the signal $\bX^*$, the data $\bY$ and of conditionally i.i.d.\ samples from the posterior $\bx^j\sim P_{X\mid Y}(\,\cdot \mid \bY)$, $j=1,2,\ldots,n$, we have that
\begin{align*}
    \EE\langle f(\bY,\bX^*,\bx^2,\ldots,\bx^{n})\rangle=\EE\langle f(\bY,\bx^1,\bx^2,\ldots,\bx^{n})\rangle,
\end{align*}
where the bracket notation $\langle \,\cdot\,\rangle$ is used for the joint expectation over the posterior samples $(\bx^j)_{j\le n}$, and $\EE$ is over the signal $\bX^*$ and data $\bY$.\\

To begin with, recall that we fixed $v$ to be the squared norm of a sample from the posterior re-scaled by the number of components. Assume that concentration effects take place, i.e. that the order parameters of the problem are limiting values of self-averaging quantities, as they should in this optimal setting \cite{barbier_strong_CMP}, and denote 
\begin{equation}
    \langle f(x)\rangle:=\frac{\int dP_X(x)e^{\sqrt{\hat{q}}Zx+\hat{m}xX-\frac{\hat{q}+\hat{v}}{2}x^2}f(x)}{\int dP_X(x)e^{\sqrt{\hat{q}}Zx+\hat{m}xX-\frac{\hat{q}+\hat{v}}{2}x^2}}.\label{eq:posterior_original}
\end{equation}
Using the Nishimori identity, we have that
\begin{equation}
    v=\lim_{N\to\infty}\frac{1}{N}\E\langle\|\bx\|^2\rangle=\lim_{N\to\infty}\frac{1}{N}\E\|\bX^*\|^2=1.
\end{equation}
We have $\hat{v}=0$ because by Bayes-optimality the constraint $v=1$ is already enforced by the prior without the need of a delta constraint. The Nishimori identity also imposes
\begin{equation}
    m=q.\label{eq:nishimori}
\end{equation}
Moreover, $B(z)$ is also fixed by the Nishimori identity (below $N$ is large and equalities are understood up to a vanishing correction as $N\to\infty$):
\begin{align}
B(z)&=\frac1N\E\langle \bx^\intercal \bG_Y(z)\bx\rangle\nonumber\\
&=\frac{1}{N}\E\bX^{*\intercal}\bG_Y(z)\bX^*\nonumber\\&=\frac{1}{N}\E\bX^{*\intercal}\Big[\bG_Z(z)+\frac{\lambda}{N}\frac{\bG_Z(z)\bX^{*}\bX^{*\intercal}\bG_Z(z)}{1-\frac{\lambda}{N}\bX^{*\intercal}\bG_Z(z)\bX^*}\Big]\bX^*\nonumber\\
&=g_Z(z)+\frac{\lambda g_Z(z)^2}{1-\lambda g_Z(z)}\nonumber\\
&=\frac{g_Z(z)}{1-\lambda g_Z(z)},
\label{eq:B(z)}
\end{align}
where we used the Nishimori identity in the second equality, Sherman-Morrison in the third one, and $$\frac{1}{N}\E\bX^{*\intercal}\bG_Z(z)\bX^*= g_Z(z)$$ in the fourth (by independence of the signal and noise).

We now state the complete set of saddle point equations obtained by cancelling the gradient of the replica free entropy potential $\{\cdots\}$ in \eqref{eq:variation_free_energy} w.r.t.\ the order parameters. The parameter w.r.t.\ which the derivative is computed in order to obtain a certain saddle point equation is reported in the round parenthesis. Let 
\begin{align}
H(x)&:=(\bar v-\bar q+2\tilde{B}(x))^{-1}, \\
R(x)&:=\bar{q}-v(\bar{m}+\tilde{M}(x))^2.    
\end{align}
Let $D\sim \rho_Z$ be drawn from the spectral distribution of the noise. Then the saddle point equations read
\begin{align*}
&(\hat{m}):\ m=\mathbb{E}X\langle x\rangle\\
&(\hat{q}):\ q=\mathbb{E}\langle x\rangle^2\\
&(\hat{v}):\ v=\mathbb{E}\langle x^2\rangle=1\\
&(\bar{m}):\ m=-\mathbb{E}_D(\bar{m}+\hat{M}(D))H(D)\\
&(\bar{q}):\ q=\mathbb{E}_DH(D)^2R(D)\\
&(\bar{v}):\ v=\mathbb{E}_DH(D)(1-H(D)R(D))\\
&(m):\ -\hat{m}+\bar{m}+\frac{m}{1-q}=0\\
&(q):\ \hat{q}-\bar{q}=\frac{q}{1-q}\\
&(v):\ \bar{v}=1\\
&(\hat{B}):\ B(z)-\frac{\lambda M(z)^2}{1-\lambda g_Z(z)}\\&\qquad\qquad=-\mathbb{E}_D\Big[\frac{1}{D-z}(H(D)-R(D)H(D)^2)\Big]\\
&(\hat{M}):\ M(z)=\mathbb{E}_D\Big[\frac{1}{D-z}(\bar{m}+\tilde{M}(D))H(D)\Big]\\
&(B):\ \frac{\lambda V'(z)}{1+\lambda B(z)}+2\hat{B}(z)=0\\
&(M):\ -\frac{2\lambda\hat{B}(z)M(z)}{1-\lambda g_Z(z)}+\hat{M}(z)=0,
\end{align*}
where we used $v=1$, $\hat{v}=0$.

We can now simplify. Firstly, from $(\hat m)$, $(\hat q)$ and \eqref{eq:nishimori}, we have
\begin{equation}
    m=q=\mathbb{E}X\langle x\rangle, \quad \text{and} \quad \hat m=\hat q.
\end{equation}
Then, from $(m)$ and $(q)$, we have
\begin{equation}
    \bar{q}=\bar{m}=\hat{m}-\frac{m}{1-m}.\label{eq:bar m}
\end{equation}
From $(\bar q)$ and $(\bar v)$, we have
\begin{equation}
    m=1-\mathbb{E}_DH(D).\label{eq:m}
\end{equation}
From $(B)$ and $(M)$, we have
\begin{align}
\hat{B}(z)&=-\frac{\lambda}{2}V'(z)(1-\lambda g_Z(z)),\label{eq:hatB}\\
\hat{M}(z)&=-\lambda^2V'(z)M(z).\label{eq:hatM}
\end{align}
Let us keep in mind that $\hat{B}(z)$ is not holomorphic, and $\hat{M}(z)$ is in general not holomorphic either.

The complex integrals in $\tilde B$ and $\tilde M$ can be performed by the residue theorem: 
\begin{align}
\tilde{B}(x)&=\frac{1}{2\pi i}\oint_\Gamma\frac{dz}{z-x}\Big[-\frac{\lambda}{2}V'(z)(1-\lambda g_Z(z))\Big]\nonumber\\
&=-\frac{\lambda}{2}V'(x)+\frac{1}{4\pi i}\oint_\Gamma\frac{dz}{z-x}\Big[\lambda^2\mathbb{E}_D\frac{V'(z)}{D-z}\Big]\nonumber\\
&=-\frac{\lambda}{2}V'(x)+\frac{\lambda^2}{2}\mathbb{E}_{D}\frac{V'(x)-V'(D)}{x-D}\label{eq:tildeB},
\end{align}
and similarly for the other function
\begin{align}
\tilde{M}(x)&=-\frac{1}{2\pi i}\oint_\Gamma\frac{dz}{z-x}\lambda^2V'(z)M(z)\nonumber\\
&=\frac{1}{2\pi i}\oint_\Gamma\frac{dz V'(z)}{z-x}\mathbb{E}\Big[\frac{\lambda^2}{z-D}(\bar{m}+\tilde{M}(D))H(D)\Big]\nonumber\\
&=\lambda^2\mathbb{E}_D\frac{V'(x)-V'(D)}{x-D}(\bar{m}+\tilde{M}(D))H(D)\label{eq:tildeM},
\end{align}
where we used $(\hat{M})$, \eqref{eq:hatB} and \eqref{eq:hatM}. 

Finally, let us denote
\begin{equation}
    J(x):=-2\tilde{B}(x),
\end{equation}
which leads to \eqref{eq:J} according to \eqref{eq:tildeB}. Recall that $\bar{v}=1$ according to $(v)$. Then, the definition of $H(x)$ and \eqref{eq:bar m} lead to \eqref{eq:H}. Combining \eqref{eq:H} with \eqref{eq:m} gives the crucial formula for the signal-to-noise ratio $\hat m$ of the effective Gaussian scalar channel associated with the model:
\begin{equation}
    \hat{m}=-\mathcal{R}_{J(\bZ)} (1-m).
\end{equation}
This forms the fixed point equations together with
\begin{equation}
    m=\mathbb{E}X\langle x\rangle_{\hat{m}},
\end{equation}
after noticing that the posterior mean \eqref{eq:posterior_original} simplifies to \eqref{eq:posterior} when $\hat{q}=\hat{m}$ and $\hat{v}=0$. The above analysis gives the simplified saddle point equations \eqref{eq:saddle_point1}. 

To obtain \eqref{eq:simiplified_f}, we simply represent all order parameters through $m$, $\hat m$ and $Q(x):=\bar m+\tilde{M}(x)$, while \eqref{eq:Q} is obtained from \eqref{eq:tildeM}. We also simplify two contour integrals as follows. The first integral is
\begin{align*}
    &-\frac{1}{2\pi i}\oint_\Gamma dz\hat M(z)M(z)\nonumber\\
    &=\frac{1}{2\pi i}\oint_\Gamma dz\lambda^2V'(z)M(z)^2\nonumber\\
    &=\frac{1}{2\pi i}\oint_\Gamma dz\lambda^2V'(z)\mathbb{E}_{D,D'}\frac{Q(D)Q(D')H(D)H(D')}{(D-z)(D'-z)}\nonumber\\
&=\lambda^2\mathbb{E}Q(D)Q(D')H(D)H(D')\frac{V'(D)-V'(D')}{D-D'},
\end{align*}
with $D,D'$ i.i.d.\ from $\rho_Z$ and where we have used \eqref{eq:hatM}, $(\hat M)$ and the residue theorem. The second integral is
\begin{align}
    \frac{1}{2\pi i}\oint_{\Gamma}dz\hat{B}(z)B(z)&=-\frac{\lambda}{4\pi i}\oint_\Gamma dzV'(z)g_Z(z)\nonumber\\&=-\frac{\lambda}{2}\mathbb{E}_DV'(D),
\end{align}
where we have used \eqref{eq:hatB} and the residue theorem.

Finally, notice that at the saddle point, specifically using \eqref{eq:B(z)}, the first term in the free entropy \eqref{eq:variation_free_energy} is precisely
\begin{align}
    -\frac{1}{4\pi i}\oint_\Gamma dz V'(z)\log (1-\lambda g_Z(z)),
\end{align}which cancels with the constant evaluated in \eqref{eq:C_final}.

\subsection{Relation to a Gaussian surrogate model} \label{sec:GaussEq}
We can notice that the replica saddle point equations (as well as the TAP equations defined in the next section) are closely related to those appearing in the Gaussian noise case. In fact, the replica saddle point equations for Gaussian noise (with SNR $\tilde{\lambda}$) read as follows:
\begin{equation}
    \hat{m}=\tilde{\lambda}^2m,\qquad
    m=\mathbb{E}X\langle x\rangle_{\hat{m}}.
\end{equation}
Therefore, by choosing $$\tilde{\lambda}=\sqrt{-\mathcal{R}_{J(\bZ)} (1-m_*)/m_*},$$ where $m_*$ takes the value at the extremizer of \eqref{eq:saddle_point1}, the Gaussian model and the rotational invariant model share the same fixed point and, thus, the same minimum mean-square error (but not necessarily the same mutual information).

\section{Thouless-Anderson-Palmer\\ free entropy and equations}
Along the lines of \cite{PNAS-Barbier-Camilli}, we employ here the adaTAP approach \cite{adatap} as an alternative to the replica method. AdaTAP offers the advantage of expressing the free entropy as a variational principle over an extensive number of parameters, which can be interpreted as site marginal means and variances for every variable in the system, namely signal components in our setting. Strictly speaking, before our work, the validity of this approach was verified only for spin glass models containing at most two-body interactions, mediated by rotationally invariant matrices. In contrast, our model is not quadratic, but the precise point of the first steps of the replica computation is to make it quadratic. Hence, we take again as a starting point (recall that the notation $\E_\bx$ means integration against the prior $P_X^{\otimes N}$):
\begin{align}
    &\Omega=\E_\bx\exp\Big(\frac{N}{4\pi i}\oint_\Gamma dz V'(z)\log(1+\frac{\lambda}{N}\bx^\intercal \bG_Y(z)\bx)\Big)\nonumber\\
    &=\E_\bx \int \mathcal{D}[B,i\hat B] \exp\Big(\frac{N}{4\pi i}\oint_\Gamma dz V'(z)\log(1+\lambda B(z))\Big)\nonumber\\
    &\qquad\times \exp\Big(\oint_\Gamma \hat B(z)(NB(z)-\bx^\intercal \bG_Y(z)\bx)dz\Big)\nonumber\\
    &=\int \mathcal{D}[B,i\hat B] \exp\Big(\frac{N}{4\pi i}\oint_\Gamma dz V'(z)\log(1+\lambda B(z))\nonumber\\
    &\qquad+N\oint_\Gamma \hat B(z)B(z) dz\Big) \E_\bx\exp\Big(\frac{1}{2}\bx^\intercal J(\bY)\bx\Big)\label{eq:quardratic_Z},
\end{align}
where $$J(\bY)=-2\oint_\Gamma \hat B(z) \bG_{Y}(z) dz$$ will end up being equal to \eqref{eq:J}. The last factor in \eqref{eq:quardratic_Z} is precisely a two-body (quadratic) model, whose partition function is therefore computable via the adaTAP approach. Define 
\begin{align}
    \varphi(\bY):=\log \E_\bx\exp\Big(\frac{1}{2}\bx^\intercal J(\bY)\bx\Big).
\end{align} Then, following Opper and Winther's prescription \cite{adatap}, the TAP representation of the auxiliary free entropy $\varphi(\bY)$ reads
\begin{align}
    \varphi_{\rm TAP}(\bY)&=\sum_{i=1}^N\Big[\lambda_i m_i+\frac{\gamma_i}{2}(\sigma_i+m_i^2)+\frac{c_i\sigma_i-\log\sigma_i-1}{2}\nonumber\\
    &\nonumber+\log\int dP_X(x)e^{-\lambda_i x-\frac{\gamma_i}{2}x^2}\Big]+\frac{1}{2}\bmm^\intercal J(\bY)\bmm\nonumber\\
    &-\frac{1}{2}\log\det(\text{diag}(\mathbf{c})-J(\bY)),
\end{align}where $\bmm=(m_i)_{i\leq N}, \mathbf{c}=(c_i)_{i\leq N}$ and an implicit extremization w.r.t.\ the parameters $\lambda_i, m_i, \gamma_i,\sigma_i,c_i$ is intended. Since we are interested only in leading terms, we can carry out some simplifications of the above. 

First of all, a common assumption in the thermodynamic limit (see \cite{adatap,maillard2019hightemp}) is that of homogeneous variances $\sigma_i=\sigma$ together with $\gamma_i=\gamma$, which in turn yields $c_i=c$. Let us now focus on the determinant term in $ \varphi_{\rm TAP}$, which is supposed to reconstruct the \emph{Onsager reaction term} in the TAP equations. We argue that at leading order it does not depend on the spike, nor on the specific realization of $\bZ$. The leading contribution is determined just by the spectral distribution of $\bZ$. Assume $J(\bY)$ is a regular enough non-linearity applied to the eigenvalues of a matrix whose spectrum consists of a bulk of eigenvalues inherited by $\bZ$, plus possibly one spike detached from the mentioned bulk. The non-linearity changes the shape of the spectrum, but it preserves the bulk-plus-spike structure. A spike of one or few eigenvalues cannot alter the spectral distribution of the overall matrix. From these considerations we get
\begin{align}
     \frac{1}{2}\log\det(\text{diag}(\mathbf{c})-J(\bY))&\simeq \frac{1}{2}\log\det(cI_N-J(\bZ))\nonumber\\
     &\simeq \frac{N}{2}\EE\log(c-J(D)),
\end{align}
where $\EE$ is intended over $D$, distributed according to the asymptotic spectral density of the noise. Hence, the TAP representation of the overall extensive free entropy of the model at leading order in $N$ reads (equalities are up to a constant and a $O_N(1)$ correction):
\begin{align}\label{eq:TAP_free_entropy}
     \log \Omega&={\rm extr}\Big\{\sum_{i=1}^N\Big[
    \lambda_i m_i+\frac{\gamma}{2}(\sigma+m_i^2)+\frac{c\sigma-\log\sigma-1}{2}\nonumber\\
    &+\log\int dP_X(x)e^{-\lambda_i x-\frac{\gamma}{2}x^2}\Big]
    -\frac{N}{2}\EE\log(c-J(D))\nonumber\\
    &+\frac{1}{2}\bmm^\intercal J(\bY)\bmm+\frac{N}{4\pi i}\oint_\Gamma dzV'(z)\log(1+\lambda B(z))\nonumber\\
    &+N\oint_\Gamma\hat{B}(z)B(z)dz\Big\}.
\end{align}
Extremization is now intended w.r.t.\ $\lambda_i,m_i,\gamma,\sigma,c$ but also the two functions $\hat B, B$. As anticipated, extremizing w.r.t. $B$ and $\hat B$ only results in matching the coupling matrix $J(\bY)$ with the pre-processed matrix using \eqref{eq:J}.

%Recall that the pre-processing function $J$ depends on the function $\hat B$, which can be determined through the $(B)$ stationarity condition:
%\begin{align}
  %  \hat B(z)=-\frac{\lambda }{4\pi i}\frac{V'(z)}{1+\lambda B(z)}.
%\end{align}
%Recalling the physical meaning of $B(z)=\frac{1}{N}\bx^\intercal \bG_{Y}(z) \bx$, and assuming concentration of the relevant order parameters such as $B$ in the thermodynamic limit, we can infer its equilibirum (average) value via Nishimori identities and the Sherman-Morrison formula. This gives \eqref{eq:B(z)},
% :
% \begin{align}
% \begin{split}
%     B(z)&\simeq\frac{1}{N}\EE\langle\bx^\intercal \bG_{Y}(z)\bx\rangle=\frac{1}{N}\EE\bX^{*\intercal} \bG_{Y}(z)\bX^*\\
%     &=\EE\frac{1}{N}(X^*)^T[G_\bZ(z)+\frac{\lambda}{N}\frac{G_\bZ(z)\bX^*(\bX^*)^\intercal G_\bZ(z)}{1-\frac{\lambda}{N}(\bX^*)^\intercal G_\bZ(z)\bX^*}]\bX^*\\
%     &\simeq g(z)+\frac{\lambda g(z)^2}{1-\lambda g_Z(z)}=\frac{g(z)}{1-\lambda g_Z(z)}\,,
% \end{split}
% \end{align}
%and hence 
%\begin{align}
%\hat B(z)=-\frac{\lambda V'(z)(1-\lambda g_Z(z))}{4\pi i}    
%\end{align}
%in the limit. Therefore, using the residue theorem, one readily gets:
%\begin{align}
%        J(\bY)&=\frac{\lambda }{2\pi i}\oint_\Gamma \bG_{Y}(z)V'(z)(1-\lambda g_Z(z))\nonumber\\
%        &=\lambda V'(\bY)-\lambda^2\mathbb{E}_D\frac{V'(\bY)-V'(D)I_N}{\bY-DI_N},
%\end{align}
%having exploited the fact that the integrand has simple poles at the eigenvalues of $\bY$, through $\bG_{Y}$, and of $\bZ$, through $g(z)$, and $D\sim\rho_Z$.

We can now write the TAP equations. Define for future convenience the Bayes ``denoiser''
\begin{align}\label{eq:denoiser}
    \eta(a,b):=\frac{\int dP_X(x)e^{ax-\frac{bx^2}{2}}\,x}{\int dP_X(y)e^{ay-\frac{by^2}{2}}}.
\end{align}
Extremization w.r.t.\ $c$ yields $\sigma=g_{J(\bZ)}(c)$, namely
\begin{align}
    c=\frac{1}{\sigma}+\mathcal{R}_{J(\bZ)} (\sigma).
\end{align}
Recall that we are looking for equilibrium configurations that satisfy Nishimori identities, so in the limit we must have
\begin{align}
    \frac{1}{N}\sum_{i=1}^N(\sigma+m_i^2)=1\text{, that is }\sigma=1-\tilde q,
\end{align}where $\tilde q:=\frac{1}{N}\sum_{i=1}^Nm_i^2$. Cancelling the $\sigma$-derivative one then gets
\begin{align}
    \gamma=-c+\frac{1}{\sigma}=-\mathcal{R}_{J(\bZ)} (\sigma)=-\mathcal{R}_{J(\bZ)} (1-\tilde q).
\end{align}
Finally, extremizing w.r.t.\ $\lambda_i$ and $m_i$ yields the final TAP equations for our model:
%\begin{align}
%  {\rm TAP \ equations}\mbox{:}\quad 
%\begin{cases}  \bmm&=\eta(J(\bY)\bmm+\gamma\bmm,\gamma),\\
   %\gamma&=-\mathcal{R}_{J(\bZ)} (1-\bar q),
   %\end{cases}
%\end{align}
\begin{align} \bmm&=\eta(J(\bY)\bmm+\gamma\bmm,\gamma),\
   \gamma=-\mathcal{R}_{J(\bZ)} (1-\bar q),
\end{align}
where $\eta$ is applied component-wise to the vector in the first entry.

The outcome of this analysis is a fundamental equivalence between the original model with non-linear likelihood governed by $V$ and a model quadratic in $\bx$, with effective interaction matrix $J(\bY)$. The equivalence is information-theoretic: the two models have asymptotically the same free entropy and, therefore, mutual information and minimum mean-square error. The main advantage of this equivalence resides in the fact that since the effective model is quadratic, we are able to employ known analytical and algorithmic approaches in the next section.

\begin{widetext}
\begin{algorithm}[Optimal data pre-processing, and TAP iterations]\label{result2}
    Define the optimal pre-processing function $J(x)$ as in \eqref{eq:J}. Let $\bmm^0=\sqrt{N}v_1(\bY)$ with $v_1(\bY)$ the unit norm first principal component of $\bY$ (or possibly another choice of initialisation). For $t\ge 1$ the TAP iterations are defined as
    \begin{align}\label{eq:TAP-iterations}
        \bmm^{t+1}=\tau\bmm^t+(1-\tau)\eta(J(\bY)\bmm^t+\gamma^t\bmm^{t-1},\gamma^t),\quad\gamma^t=-\mathcal{R}_{J(\bZ)} (1-\tilde q^t),\quad \tilde q^t=\frac{\|\bmm^t\|^2}{N},
    \end{align}
    where $\eta$ is defined in \eqref{eq:denoiser}, we use a damping parameter $\tau\in[0,1)$ for improved numerical stability, and $\mathcal{R}_{J(\bZ)} $ is the $R$-transform of the asymptotic spectral density of $J(\bZ)$. 
\end{algorithm} 
\end{widetext}

\section{From TAP equations to an efficient iterative algorithm}

Now that the information-theoretic analysis has been performed through the replica method, we switch towards algorithmic aspects and focus on how to efficiently match the performance predicted by our theory, based on the Thouless-Anderson-Palmer formalism \cite{thouless1977solution,MezardParisi87b,adatap}.

\subsection{TAP iterations}
%\jb{explain here how we go from TAP to the algo state here: time indices inspired by AMP, spectral initialization through a power method, complexity}

We can now state our second main result ({\bf Algorithm} above), which is of an algorithmic nature. This comes in the form of a Bayes-optimal pre-processing function to be applied to the data matrix, and an efficient iterative algorithm exploiting it, in order to reach a solution of the TAP equations.

We draw attention to the time indexing in the algorithm \eqref{eq:TAP-iterations}. The update rule is inspired by that of a usual AMP algorithm, and as supported by our numerical experiments, it proves to be effective to match the results predicted by the replica analysis. Despite its similarity, with an evident candidate Onsager reaction term, $\gamma^t \bmm^{t-1}$, \eqref{eq:TAP-iterations} cannot be really regarded as an AMP algorithm per se, since we have no theoretical guarantee that the components of the iterates $J(\bY)\bmm^t+\gamma^t\bmm^{t-1}$ have empirical Gaussian statistics. 
%Thanks to the low degree of the polynomial potential selected in the work \cite{PNAS-Barbier-Camilli}, in that occasion it was possible to find the correct Onsager reactions and an AMP algorithm with a state evolution. Having a generic non-quadratic potential however, has prevented us from finding Onsager reactions in a closed form so far. Nevertheless, the adaTAP approach gives a compact, physically interpretable, and potentially Bayes-optimal effective alternative algorithm. We have  investigated numerically the properties of the fixed points of \eqref{eq:TAP-iterations}, and the results are summarized in the next subsection.

\begin{figure*}[p]
    \centering
    {
        \includegraphics[width=0.48\linewidth]{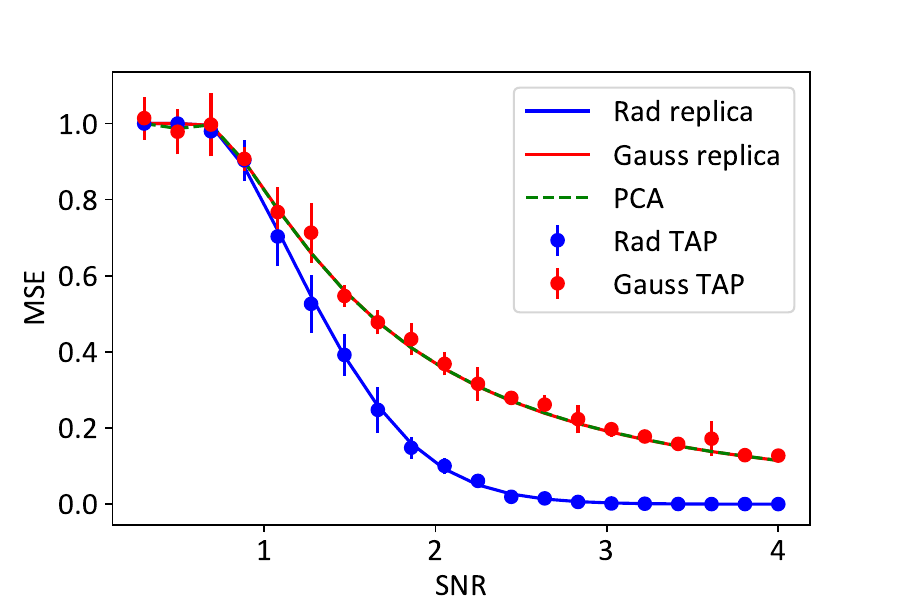}
        \includegraphics[width=0.48\linewidth]{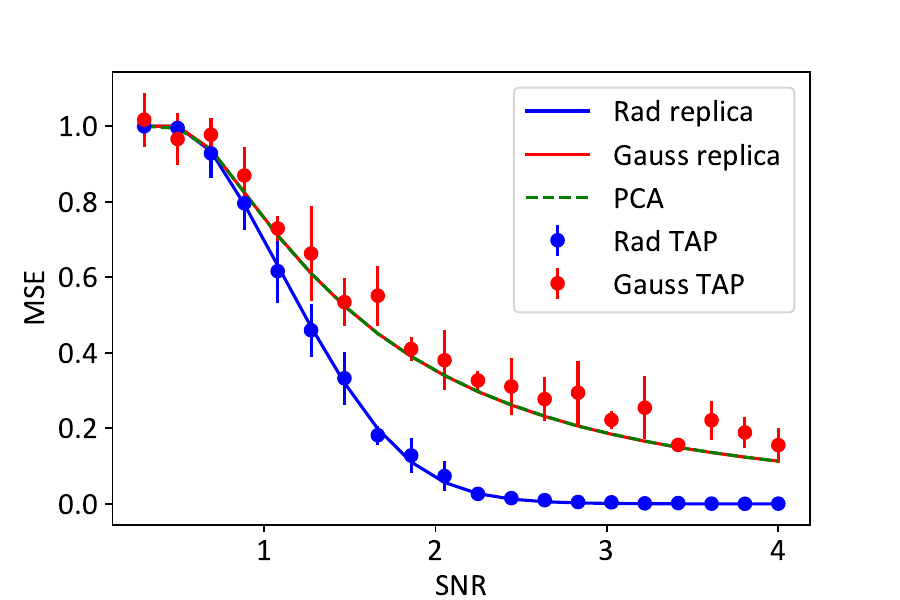}\\
        \includegraphics[width=0.48\linewidth]{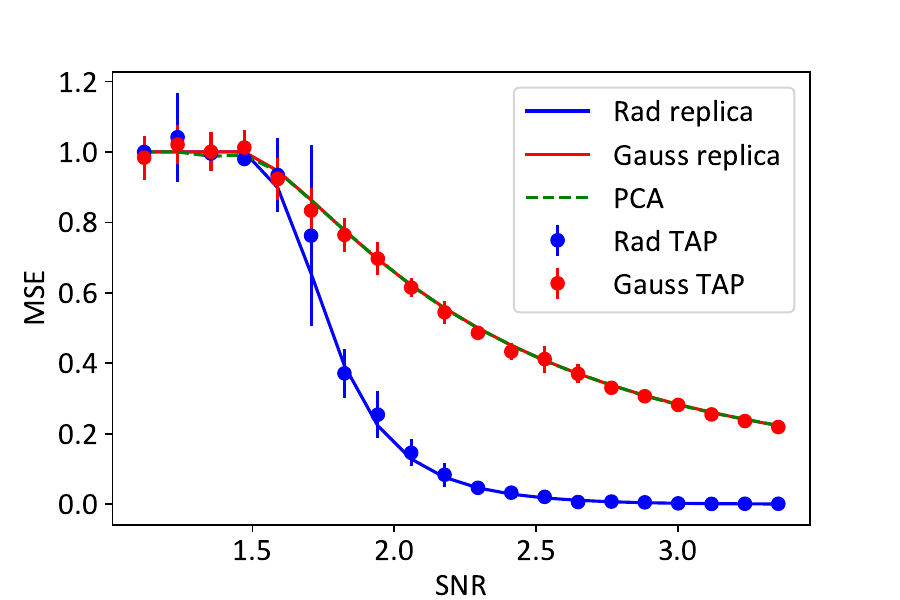}
        \includegraphics[width=0.48\linewidth]{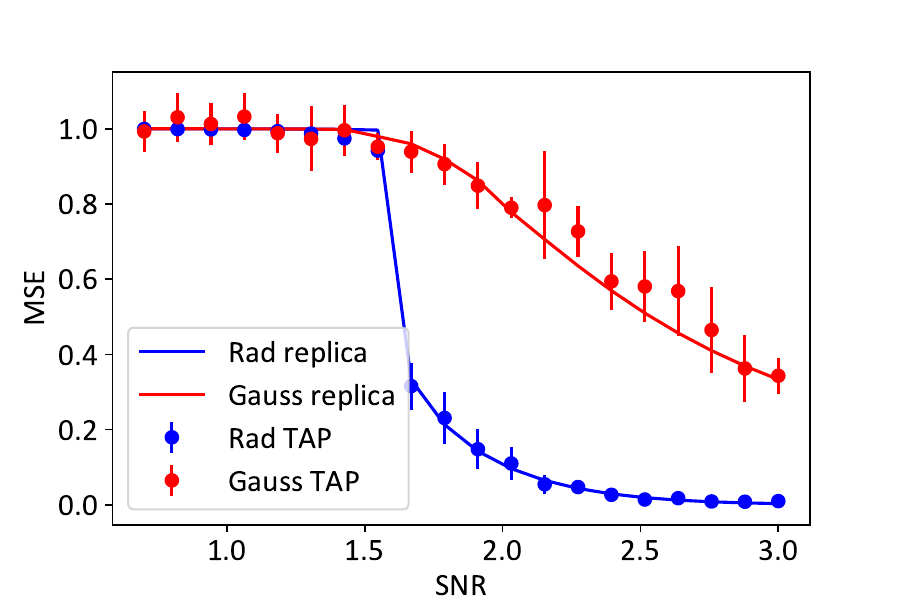}
    }
    \caption{The performance of the TAP iterations (dots) matches well the replica prediction for the minimum mean-square error (solid lines), for various distributions of noise eigenvalues (in different plots) and for two signal priors (Gaussian in red and Rademacher in blue). Error bars correspond to the standard deviation over 10 trials. \textbf{Top left}: Quartic potential. \textbf{Top right}: Sestic potential. \textbf{Bottom left}: Marchenko–Pastur distribution of eigenvalues. \textbf{Bottom right}: Truncated normal distribution of eigenvalues. The green dashed lines (which overlap perfectly the red solid lines) denote the theoretical performance of spectral PCA as predicted by \cite{benaych2011eigenvalues}. We do not include the performance of spectral PCA for the normal distribution of eigenvalues  due to numerical instabilities.}
\label{fig:four_cases}
\end{figure*}

\begin{figure*}[p]
    \centering
    {
        \includegraphics[width=0.48\linewidth]{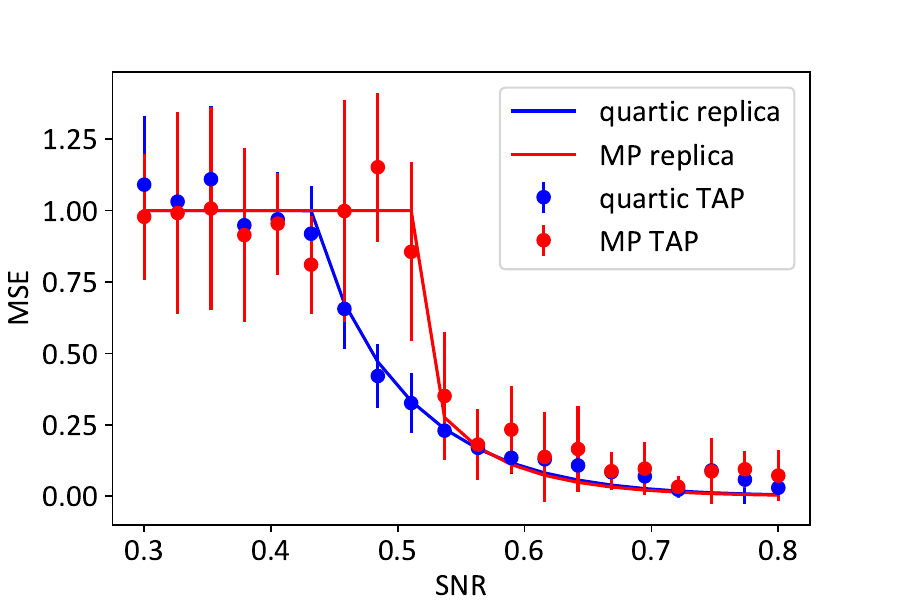}
        \includegraphics[width=0.48\linewidth]{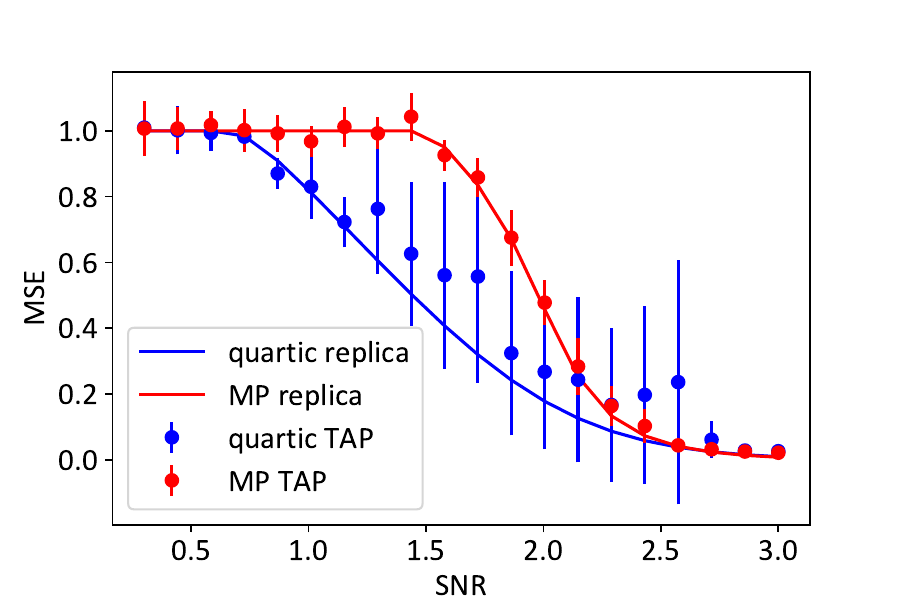}
    }
    \caption{The performance of the TAP iterations (dots) matches well the replica prediction for the minimum mean-square error (solid lines), for two sparse priors (in different plots) and two distributions of noise eigenvalues (quartic potential in red and Marchenko–Pastur distribution in blue). Error bars correspond to the standard deviation over 10 trials. \textbf{Left}: two-point prior. \textbf{Right}: sparse Rademacher prior.}
\label{fig:sparse_cases}
\end{figure*}

\subsection{Numerical experiments}\label{sec:numerical_experiments}
We refer the reader to \cite{Note1} for the code related to this section.

To verify the match between our replica theory and algorithm, we choose four concrete examples for the noise potential. $(i)$ A quartic potential $V(x)=\gamma x^4/4$ with $\gamma=16/{27}$. Its eigenvalue distribution is given by
\begin{equation}
    \rho_Z(x)=\frac{1}{2\pi}(2a^2\gamma+\gamma x^2)\sqrt{4a^2-x^2},
\end{equation}
where $a=3/4$. $(ii)$ A sestic potential $V(x)=\xi x^6/6$ with $\xi=27/80$. Its eigenvalue distribution is given by
\begin{equation}
    \rho_Z(x)=\frac{1}{2\pi}(6a^4\xi+2a^2\xi x^2+\xi x^4)\sqrt{4a^2-x^2},
\end{equation}
where $a=\sqrt{2/3}$. In both cases, the constants are chosen in order to enforce unit variance for the spectral densities. These two cases are (among) those studied in the previous paper \cite{PNAS-Barbier-Camilli}, the sestic potential being the highest degree of a polynomial that the techniques in the reference allowed to study. With the present contribution we can now analyse arbitrary potentials, even non-polynomial ones such as $(iii)$ eigenvalues following the Marchenko–Pastur distribution:
\begin{equation}
    \rho_Z(x)=\frac{1}{2\pi\sigma^2}\frac{\sqrt{(\lambda_+-x)(x-\lambda_-)}}{\alpha x},\label{eq:MP_distribution}
\end{equation}
where $\lambda_\pm=\sigma^2(1\pm\sqrt{\alpha})^2$ and $\alpha=0.2$. The associated potential is given by $V(x)=[(1-1/\alpha)/x+1/\alpha]/\sigma^2$. Finally, %as an example of more exotic distribution of eigenvalues 
we consider $(iv)$ eigenvalues following a standard normal distribution truncated between $[-5,5]$. Its potential has probably no analytical expression, so we numerically calculated its derivative through \cite{potters2020first}
\begin{equation}
    V'(x)=2\,\text{P.V.}\int\frac{\rho_Z(d\lambda)}{x-\lambda}\label{eq:V'},
\end{equation}
where $\text{P.V.}$ denotes the Cauchy principal value. Thus, we are able to calculate $J(\bZ)$ and its R-transform. In all cases, the noise is properly normalized such that $\mathbb{E}(D-\mathbb{E}D)^2=1$, which is also how we determine $\sigma^2$ in \eqref{eq:MP_distribution}. We consider four choices for the prior $P_X$: \emph{(i)} a standard Gaussian prior $\mathcal{N}(0,1)$, \emph{(ii)} a Rademacher prior $\frac12\delta_{-1}+\frac12\delta_{1}$, \emph{(iii)} a two-point prior $\epsilon^2\delta_{1/\epsilon}+(1-\epsilon^2)\delta_0$ with $\epsilon=0.125$, and \emph{(iv)} a sparse Rademacher prior $(1-\epsilon^2)\delta_0+\frac{\epsilon^2}{2}(\delta_{-1/\epsilon}+\delta_{-1/\epsilon})$ with $\epsilon=\sqrt{0.3}$.

The algorithm uses a PCA initialization \cite{mondelli2021pca,zhong2021approximate}, that can be obtained efficiently via the power method. For the normally distributed eigenvalues and two-point prior signals, however, we manually choose an initialization having positive correlation $\sqrt{0.5}$ with the ground truth for numerical stability. In all experiments, we use $N=2000$ and show the results averaged over $10$ trials and the corresponding standard deviation. In some cases, about $20\%$ of the trials does not converge to the right fixed point, which we exclude when gathering statistics. Moreover, we fix the Onsager coefficient to its fixed point value predicted by the replica theory for numerical stability. We use a damping $\tau=0.9$ in all experiments. 

Fig.~\ref{fig:four_cases} and Fig.~\ref{fig:sparse_cases} show that in all successful cases, our algorithm approaches the Bayes-optimal performance predicted by our replica-based theory. We therefore conjecture that the fixed point performance of our algorithm matches that of the minimum mean-square error estimator, when no statistical-to-computational gap is present. %, as in all the above experiments. 

In Fig.~\ref{fig:four_cases}, we also report the performance corresponding to the spectral PCA initialization alone, as predicted by \cite{benaych2011eigenvalues}. As noticed in \cite{PNAS-Barbier-Camilli} for low degree polynomial potentials, PCA remains Bayes-optimal when the prior of the signal is Gaussian (or more generically rotationally invariant), regardless of the noise eigenvalue distribution.

We further note that, as can be seen from Fig.\ \ref{fig:four_cases}, a Rademacher prior leads to better numerical stability, due to a more attractive fixed point for the dynamics. In fact, the Rademacher prior -- being more informative -- constrains the signal estimate more strongly than a Gaussian prior. 

\begin{figure}[t!]
    {
        \includegraphics[width=\linewidth]{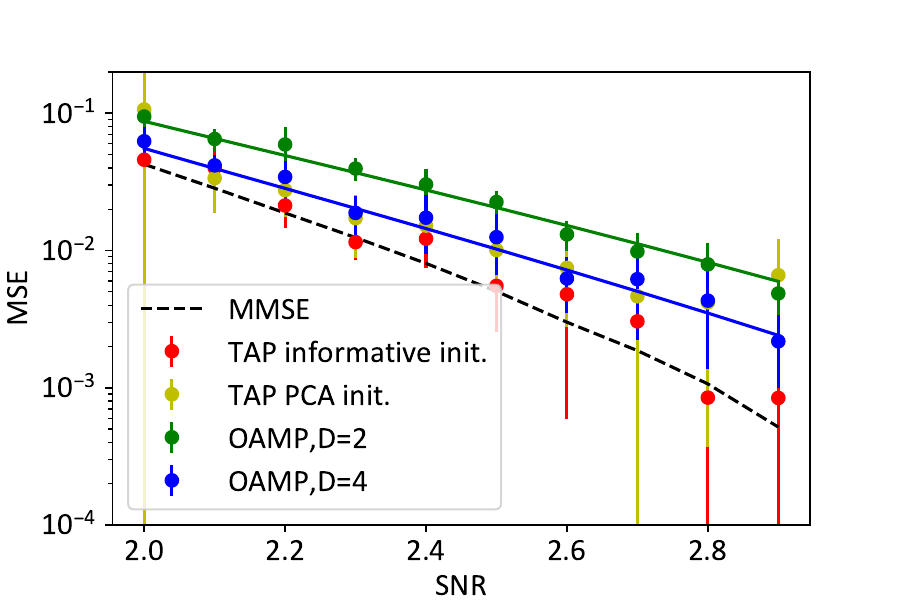}
    }
    \caption{Comparison between the TAP algorithm and the optimal degree-D lifted OAMP algorithm \cite{dudeja2024-OAMP} on noise matrices derived from the Hapmap3 dataset and Rademacher signals. Solid lines correspond to the state evolution of the optimal degree-D lifted OAMP algorithm and the dashed black line is the replica prediction for the MMSE.}
\label{fig:hapmap}
\end{figure}

\begin{figure}[t!]
    {
        \includegraphics[width=\linewidth]{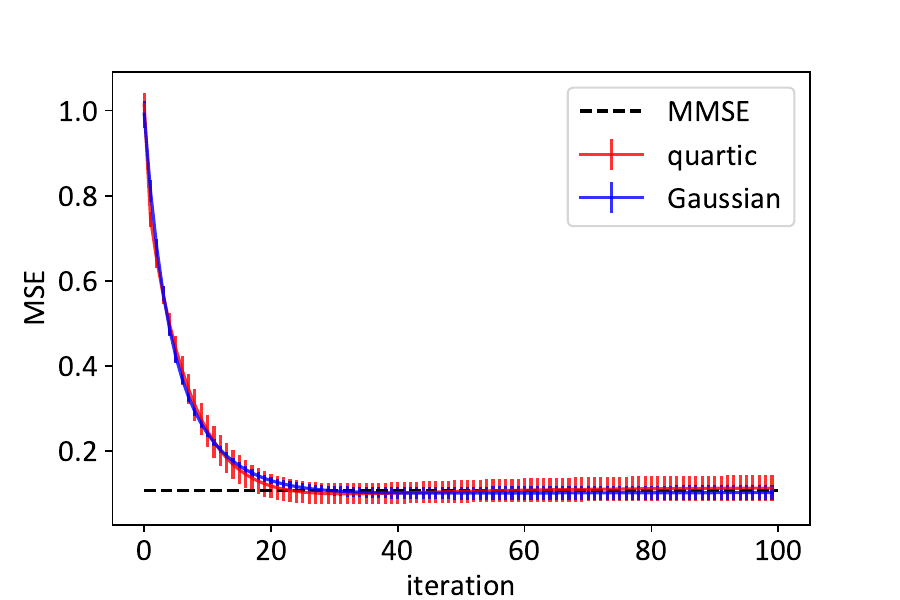}
    }
    \caption{Comparison between the TAP iterations for the quartic noise model (with $\lambda=2$) and its information-theoretic equivalent Gaussian surrogate model. The error bars represent the standard deviation computed over 10 trials. The dashed black line represents the MMSE predicted by the replica theory.}
\label{fig:ite}
\end{figure}

Next, we test our TAP algorithm on a popular dataset in bioinformatics -- the International HapMap Project (Hapmap3) \cite{international2010integrating}, which comprises $142186$ single nucleotide polymorphisms (SNPs) for $1397$ individuals. In our experiments, we randomly select $3000$ SNPs and compute the corresponding covariance matrix twice. We calculate the difference of the two sample covariance matrices as the noise matrix. Then, we extract $8$ principal components that correspond to
outlying eigenvalues. We also center and scale the matrix such that the eigenvalues have zero mean and unit variance. We then plant a spike constructed from a random vector with Rademacher law.

We compare with the approach of the concurrent paper \cite{dudeja2024-OAMP} that focuses on the same setting and  develops a new class of AMP algorithms, together with a rigorous state evolution result for them. A fixed point of that AMP is shown to match the replica predictions of \cite{PNAS-Barbier-Camilli} and, in fact, we verify that such a fixed point matches the replica predictions we make in the present paper as well (see the Appendix). Therefore, the algorithm of \cite{dudeja2024-OAMP}, just as the one we propose here, is conjectured to be Bayes-optimal when no statistical-to-computational gap is present.

In Fig.~\ref{fig:hapmap}, we report the performance of the optimal degree-D lifted OAMP algorithm \cite{dudeja2024-OAMP} with PCA initialization, as well as that of our TAP algorithm with PCA initialization or sufficiently informative intialization (correlation $\sqrt{0.9}$). The optimal degree-D lifted OAMP algorithm is a sub-optimal algorithm proposed in \cite{dudeja2024-OAMP} for real data because it only relies on moments of the noise eigenvalues. Here the MSE is that of the signal $\bX^*$, instead of the spike $\bX^*\bX^{*\intercal}$, because the state evolution of OAMP is for the MSE of the signal. The corresponding MMSE predicted by the theory is thus $1-m_*$. To implement TAP, we calculate the pre-processing function using the empirical eigenvalue distribution of the noise matrix. The plot shows that our TAP algorithm with sufficiently informative intialization performs closely to the replica prediction of the optimal MSE, while our TAP algorithm with PCA initialization performs similarly to the optimal degree-$4$ lifted OAMP algorithm. It is left as future work to overcome the gap between practical algorithms and replica predictions.

Finally, the equivalence between the models with structured and Gaussian noises (see sec.~\ref{sec:GaussEq}) does not only hold at the level of static (thermodynamic) properties. Indeed, Fig.~\ref{fig:ite} numerically verifies that the gap between the TAP iterations run for the Gaussian model and those run for the rotational invariant model is small.

\section*{Acknowledgments}

J.B., F.C. and Y.X. were funded by the European Union (ERC, CHORAL, project number 101039794). Views and opinions expressed are however those of the authors only and do not necessarily reflect those of the European Union or the European Research Council. Neither the European Union nor the granting authority can be held responsible for them. M.M. was supported by the 2019 Lopez-Loreta Prize. J.B. acknowledges discussions with TianQi Hou at the initial stage of the project, as well as with Antoine Bodin.

\section*{Appendix: Equivalence between the replica saddle point equations and the state evolution in \cite{dudeja2024-OAMP}} \label{app}

In \cite{dudeja2024-OAMP}, the state evolution reads (without time indices)
\begin{align}
    &\theta=\frac{1}{\text{dmmse}(\omega)}-1,\label{eq:theta}\\
    &\omega=1-\Big(\mathbb{E}_D\Big[\frac{\phi(D)}{\phi(D)+\theta}\Big]\Big)^{-1}\mathbb{E}_D\Big[\frac{1}{\phi(D)+\theta}\Big]\label{eq:omega},
\end{align}
where $D\sim\rho_Z$ as in the main text and
\begin{equation}
    \phi(x):=(1-\pi\lambda \mathcal{H}(x) )^2+\pi^2\lambda^2\rho_Z^2(x).
\end{equation}
Here, $\mathcal{H}(x) $ denotes the Hilbert transform of $\rho_Z(x)$ and 
$\text{dmmse}(\omega)$ represents the divergence-free minimum mean-squared error (DMMSE) of a Gaussian channel with prior $P_X$ and signal-to-noise ratio $\sqrt{\omega/(1-\omega)}$. The DMMSE is related to the MMSE of the same channel through 
\begin{equation}
    \frac{1}{\text{dmmse}(\omega)}=\frac{1}{\text{mmse}(\omega)}-\frac{\omega}{1-\omega}
\end{equation}
for $\omega<1$ (\cite[Lemma 2]{dudeja2024-OAMP}).

Denote 
\begin{equation}
    m:=1-\text{mmse}(\omega),\quad \hat{m}:=\frac{\omega}{1-\omega},
\end{equation}
from which we have the first replica equation of \eqref{eq:saddle_point1}
\begin{equation}
    m=\mathbb{E}X\langle x\rangle_{\hat{m}}.
\end{equation}

From \eqref{eq:theta} we have
\begin{equation}
    \theta=\frac{1}{1-m}-\frac{1}{1-\omega}.\label{eq:A7}
\end{equation}
From \eqref{eq:omega} we have
\begin{equation}
    \mathbb{E}_D\Big[\frac{1}{\phi(D)+\theta}\Big]=\frac{1-\omega}{1+\theta(1-\omega)},
\end{equation}
leading to
\begin{equation}
    \mathbb{E}_D\Big[\frac{1}{\phi(D)+\theta}\Big]=1-m,
\end{equation}
or equivalently
\begin{equation}
    \theta-\frac{1}{1-m}=\mathcal{R}_{-\phi(\bZ)}(1-m).
\end{equation}
Using \eqref{eq:A7}, we obtain
\begin{equation}
    \hat{m}=-\mathcal{R}_{1-\phi(\bZ)}(1-m).
\end{equation}
The relation between $\phi(\cdot)$ and $J(\cdot)$ is given by
\begin{equation}
\begin{aligned}
1-\phi(x)&=2\lambda\pi \mathcal{H}(x) -\lambda^2[\pi^2\mathcal{H}^2(x)+\rho_Z^2(x)]\\&=\lambda V'(x)-\lambda^2\mathbb{E}_D\frac{V'(x)-V'(D)}{x-D},\label{eq:A12}
\end{aligned}
\end{equation}
where the last equality is due to $2\pi \mathcal{H}(x) =V'(x)$, see \eqref{eq:V'}, and \cite[Lemma 10]{dudeja2024-OAMP}. Notice that the right side of \eqref{eq:A12} is exactly $J(x)$, and thus we recover 
\begin{equation}
    \hat{m}=-\mathcal{R}_{J(\bZ)} (1-m),
\end{equation}
which is the second replica saddle point equation in \eqref{eq:saddle_point1}. 

\bibliography{biblio}% Produces the bibliography via BibTeX.

\end{document}